\documentclass[prl,twocolumn,showpacs,superscriptaddress,floatfix]{revtex4}
\usepackage{graphicx,amsfonts,amssymb,amsmath, hyperref}

\newif\ifhyper
\hypertrue
\ifhyper
\hypersetup{
   citecolor = {green},
   colorlinks = {true}, 
   urlcolor = {blue} 
}
\fi

\newcommand{\beq}{\begin{equation}}
\newcommand{\eeq}{\end{equation}}
\newcommand{\beqa}{\begin{eqnarray}}
\newcommand{\eeqa}{\end{eqnarray}}
\newcommand{\ket} [1] {\vert #1 \rangle}

\newcommand{\braket}[2]{\langle #1 | #2 \rangle}

\begin{document} 

\title{Equivalence of critical scaling laws for many-body entanglement in the Lipkin-Meshkov-Glick model}

\author{Rom\'an Or\'us}
\email{orus@physics.uq.edu.au}
\affiliation{School of Physical Sciences, The University of Queensland, QLD 4072, Australia}
 
\author{S\'ebastien Dusuel}
\email{sdusuel@gmail.com}
\affiliation{Lyc\'ee Louis Thuillier, 70 Boulevard de Saint Quentin,
  80098 Amiens Cedex 3, France}

\author{Julien Vidal}
\email{vidal@lptmc.jussieu.fr}
\affiliation{Laboratoire de Physique Th\'eorique de la Mati\`ere Condens\'ee,
  CNRS UMR 7600, Universit\'e Pierre et Marie Curie, 4 Place Jussieu, 75252
  Paris Cedex 05, France}

\begin{abstract}

We establish a relation between several entanglement properties in the Lipkin-Meshkov-Glick model, which is a system of mutually interacting spins embedded in a magnetic field.  We provide analytical proofs that the single-copy entanglement and the global geometric entanglement of the ground state close to and at criticality behave as the entanglement entropy. These results are in deep contrast to what is found in one- dimensional spin systems where these three entanglement measures behave differently.  

\end{abstract}


\maketitle

%
%
\emph{Introduction ---}
%
%
In the field of quantum many-body physics, common wisdom dictates that quantum entanglement plays a key role in the occurrence of important collective phenomena at zero temperature \cite{Sachdev99}. Understanding the entanglement properties of many-body systems at the critical points of quantum phase transitions is, therefore, a great theoretical challenge. In this respect, considerable efforts have been devoted in recent years towards a theory of entanglement in extended systems (see \cite{Amico08}  for a review). 

Within this context, significant attention has been paid to systems in one dimension (1D), which are mostly tractable by analytical studies. In particular, the pioneering works \cite{Vidal03_1,Latorre04_1,Its05} established that the ground state {\it entanglement entropy} ${\mathcal E}$  obeys universal scaling laws in critical regions (that is, either close to or at criticality) that are described by an underlying conformal field theory \cite{Holzhey94}. 
This entropy measures the entanglement between two subspaces and thus relies on a bipartition of the original system.  
Along the same line, universal scaling laws have been also obtained for the {\it single-copy entanglement} ${\mathcal S}$ \cite{Eisert05,Orus06_1} which quantifies the amount of entanglement distillable from a single specimen of a quantum  system, and for the density of {\it global geometric entanglement} per subsystem $\mathcal{G}/N$ \cite{Botero08, Orus08_1} which measures the distance, in the Hilbert space, to the closest separable state. Surprisingly enough, for critical 1D systems, these three quantities which are {\it a priori} very different,  turn out to be deeply intertwined since they obey $\mathcal{G}/N \sim {\mathcal S}/2 \sim {\mathcal E}/4$.  
This result has important consequences in our understanding of renormalization group flows in 1D \cite{Zamolodchikov86,Latorre05_1,Orus05,Zhou06}.

It is then natural to wonder whether such a relation still holds for higher-dimensional critical systems. However, answering this question implies to solve some difficult problems. Indeed, the majority of systems beyond 1D do not admit an analytical solution and their properties need to be unveiled by numerical simulations, always with a partial success \cite{Nishino98,Verstraete04_3,Jordan07,Sandvik07}. 

The aim of this letter is to investigate this issue by providing an analytical derivation of the relation at criticality between the entanglement entropy ${\mathcal E}$, the single copy entanglement ${\mathcal S}$ and the global geometric entanglement ${\mathcal G}$ in the Lipkin-Meshkov-Glick (LMG) model \cite{Lipkin65,Meshkov65,Glick65}. 

Originally introduced in nuclear physics, the LMG model has, since then,  been used in the description of many physical systems among which  are two-mode Bose-Einstein condensates \cite{Cirac98} or small ferromagnetic particles \cite{Chudnovsky88}. However, the full spectrum of this model has only been  exactly determined recently in the thermodynamical limit \cite{Ribeiro07} and has revealed a very rich structure.
Some entanglement properties have already been investigated in this model (concurrence \cite{Vidal04_1,Dusuel04_3,Dusuel05_2}, entropy\cite{Latorre05_2,Barthel06_2,Vidal07}, fidelity \cite{Kwok08}) albeit, contrary to 1D systems, there is no comparison amongst different measures. 

In this work, we bridge this gap by computing exactly, in the thermodynamical limit, the global geometric entanglement as well as the single-copy entanglement of the ground state. These results allow us to extract their behavior in the critical region and to establish that these quantities obey \emph{exactly the same scaling laws} as the entropy, {\it i.e.}, ${\mathcal G} \sim {\mathcal S} \sim {\mathcal E}$ in deep contrast with one-dimensional spin systems. 

%
%
\emph{The model ---}
%
%
The LMG model describes a system of $N$ spins $1/2$ mutually interacting and embedded in a transverse magnetic field. Its dimensionless Hamiltonian is given by
%
%
\beq
H = -\frac{1}{N} \left( S_x^2 + \gamma \: S_y^2 \right) - h S_z ,
\label{LMG}
\eeq
%
%
where $S_{\alpha} = \sum_{i=1}^N \sigma_{\alpha}^i/2$ are the total spin operators in the direction $\alpha$, 
$\sigma_{\alpha}^i$  is the Pauli matrix $\alpha$ for spin $i$, $\gamma$ is the anisotropy parameter and $h$ is the transverse magnetic field. Notice that the Hamiltonian  (\ref{LMG}) can be seen as those of a $N$-dimensional system of $N$ spins $1/2$ or a zero-dimensional system of one spin $N/2$ particle. 
Here, we focus on the ferromagnetic case and, without loss of generality, we assume $0 \le \gamma < 1$ and $h \ge 0$. 

It is well-known that the system undergoes a second-order quantum phase transition at $h = 1$ which separates a symmetric phase for $h > 1$ from a broken phase $h < 1$.  The basic properties of these phases can be understood in terms of a mean-field approach \cite{Botet83,Dusuel05_2}. Within this approximation, the ground state is described by a fully-polarized state which is unique for $h > 1$, and two-fold degenerate for $h < 1$. 
However, although the mean-field treatment perfectly describes the transition, the exact ground state is not a product state, even in the thermodynamic limit, as initially observed for the concurrence \cite{Vidal04_1}.
This counter-intuitive result can be easily understood in terms of quantum fluctuations around a classical state by using a Holstein-Primakoff representation of the spin operators \cite{Dusuel05_2}.  

In the following, for simplicity, we further restrict our study to the symmetric phase and the critical point, that is, the region $h \ge 1$.  

%
%
\emph{The global geometric entanglement ---}
%
%
To introduce this measure, let us consider a pure quantum state of $N$ parties $\ket{\Psi} \in \mathcal{H} = \bigotimes_{i = 1}^N \mathcal{H}^{[i]}$, where $\mathcal{H}^{[i]}$ is the Hilbert space of party $i$. 
We wish to quantify the global multipartite entanglement of $\ket{\Psi}$. Following \cite{Wei03}, this is achieved by considering the maximum fidelity $\Lambda_{{\rm max}}$ between $\ket{\Psi}$ and all the possible separable states $\ket{\Phi}$ of the $N$ parties 
%
%
\beq
\Lambda_{{\rm max}} = {\rm max} |\braket{\Phi}{\Psi} |.
\label{scalar}
\eeq
%
%
This quantity can be seen as the distance, in the Hilbert space, between the state $\ket{\Psi}$ and the closest separable state. Here, a state $|\Phi\rangle$ is said to be separable if it can be written as  $|\Phi\rangle= \otimes_{i=1}^N (u_i \ket{\uparrow}_i+ v_i \ket{\downarrow}_i)$ where $\ket{\uparrow}_i$ ($\ket{\downarrow}_i$) is the eigenstate of $\sigma_z^i$ with eigenvalue $+1$ ($-1$).
Thus, we assume that each individual spin constitutes a single party by contrast to the analysis done in Refs.~\cite{Botero08, Orus08_1}, where the parties are blocks of several spins.
In order to have a well-defined measure of entanglement (which vanishes when $\ket{\Psi}$ is a product state) one defines the global geometric entanglement ${\mathcal G}$ of state $\ket{\Psi}$ as
%
%
\beq
{\mathcal G}= - \ln{\Lambda_{{\rm max}}^2}.
\eeq
%
%

\begin{figure}[t]
  \includegraphics[width=\columnwidth]{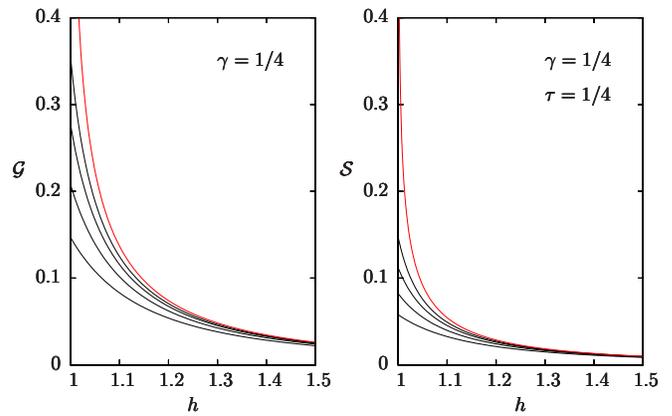}
  \caption{(color online).
Behavior of the geometric entanglement (left panel ) and single-copy
entanglement (right panel) as a function of the magnetic field for different
system sizes $N=16$, $32$, $64$, $128$ and $256$ (from bottom to top).  Red lines are the thermodynamical limit given in Eqs.~(\ref{eq:Gthermo}) and (\ref{eq:Sthermo}). Here, ${\mathcal G}$ has been obtained by a numerical minimization of the fidelity over all coherent states.}
  \label{fig:GSvsh}
\end{figure}

Notably, for the LMG model, one already knows the state $\ket{\Phi}$ which maximizes the fidelity. 
Indeed, one knows that the ground state is in the maximum spin sector $S=N/2$ which is permutation-invariant. This implies that the closest separable state in the Hilbert space must be of the form
$|\Phi\rangle= \otimes_{i=1}^N (u \ket{\uparrow}_i+ v \ket{\downarrow}_i)$. 
One thus seeks for a coherent state which is as close as possible to the exact ground state. The answer is given by the mean-field treatment detailed in Refs.~\cite{Botet83,Dusuel05_2} and, in the symmetric phase, one gets $u=1$ and $v=0$, {\it i.~e.}, the fully-polarized state  in the $z$-direction. 
The next step consists in computing $\Lambda_{{\rm max}}$ which is more involved since one does not know the exact ground state analytically. Nevertheless, in the thermodynamical limit (large $N$), as early given in the seminal paper \cite{Meshkov65}, one can obtain a recursion relation for the coefficients in the Dicke states basis. After simple algebra, one gets the following expression of the ground state 
%
%
\beq
\ket{\Psi_0} = (1-t^2)^{1/4} \sum_{i = 0}^{N/2} 
{\left(
\begin{array}{c}
2i
\\
i
\end{array}
\right)^{1/2}
} 
\left(\frac{t}{2}\right)^i \ket{2 i},
\label{state}
\eeq
%
%
where 
%
%
\beq
t =\frac{2h-\gamma-1-2\sqrt{(h-1)(h-\gamma)}}{1-\gamma}.
\eeq
%
%
Here, the state $|2i \rangle$ denotes the eigenstate of ${\bf S}^2$ and $S_z$ with eigenvalues $\frac{N}{2}(\frac{N}{2}+1)$ and $\frac{N}{2}-2 i $. The maximum fidelity is then directly given by the coefficient on the state $\ket{0}$ so that, in the thermodynamical limit, the global geometric entanglement reads
%
%
\beq
{\mathcal G}(\gamma,h) = -\frac{1}{2} \ln \big(1-t^2 \big)  .
\label{eq:Gthermo}
\eeq
%
%
As can be checked in Fig.~\ref{fig:GSvsh} (left), this is in perfect agreement with numerics. 
%
%
%
%

Near the critical point $h=1$, this quantity behaves as
%
%
\beqa
{\mathcal G}(\gamma,h)&=&-\frac{1}{4} \ln(h-1) + \frac{1}{4} \ln(1-\gamma) - \ln 2 \nonumber \\
&& + {\mathcal O} \big[(h-1)^{1/2}\big].
\label{offe0}
\eeqa
%
%
Quite importantly, the above relation shows that when reaching the critical point ${\mathcal G}$ diverges as $-\frac{1}{4} \ln(h-1)$, exactly as the entanglement entropy 
${\mathcal E}$ \cite{Barthel06_2}. We emphasize this result is completely nontrivial since ${\mathcal G}$ and ${\mathcal E}$ are, by construction, very different objects.
As we shall now see, it is even more surprising to see that the scaling laws derived above are also valid for the single-copy entanglement. 

%
%
\emph{The single-copy entanglement ---}
%
%
Let us now introduce the next entanglement measure for a quantum many-body state $\ket{\Psi}$ with reduced density matrix $\rho_L$ for a subset of $L$ constituents. As explained in Refs. \cite{Eisert05, Orus06_1}, the so-called single-copy entanglement ${\mathcal S}$ between this subset and the rest of the system is given by
\beq
{\mathcal S}=-\ln \lambda_1  ,
\eeq
%
%
where $\lambda_1$ is the largest eigenvalue of the reduced density matrix $\rho_L$. 

Our aim is thus to compute and diagonalize the reduced density matrix $\rho_L$ obtained by tracing out over $(N-L)$ spins in the LMG model. As detailed in Ref.~\cite{Dusuel05_2}, the Hamiltonian (\ref{LMG}) can be mapped, in the thermodynamical limit,  onto a quadratic form of a single bosonic mode $a$ via a standard Holstein-Primakoff transformation of the spin operators. As early shown by Bombelli {\it et~al.} \cite{Bombelli86}, one also knows that the reduced density matrix for eigenstates of a quadratic form can always be written as $\rho_L = {\mathrm e}^{-K}$ with 
%
%
\beq
K = \kappa_0 + \kappa_1 a^{\dagger} a + \kappa_2 (a^{\dagger 2} + a^2)  , 
\eeq
%
%
where the coefficients $\kappa_0$, $\kappa_1$ and $\kappa_2$ have to be determined by self-consistent relations. Following Refs.~\cite{Barthel06_2,Vidal07}, after diagonalization of $K$, one gets for the LMG model
%
%
\beq
\rho_L = \left( \frac{2}{\mu + 1} \right) {\mathrm e}^{-\epsilon \, g^{\dagger} g} ,
\eeq
%
%
where $g$ is a bosonic mode that diagonalizes $K$, and
%
%
\beq   
\mu = \alpha^{-1/2} \sqrt{[\tau \alpha + (1 - \tau)][\tau + \alpha(1-\tau)]},
\eeq
%
%
with $\tau = L/N$ and $\alpha = \sqrt{(h-1)/(h-\gamma)}$. The pseudoenergy reads $\epsilon = \ln \left( \frac{ \mu + 1}{\mu - 1} \right)$.
The single-copy entanglement of a subset of $L$ spins, in the thermodynamical limit,  is thus given by 
%
%
\beq
{\mathcal S}(\tau,\gamma,h) = \ln \frac{\mu + 1}{2}. 
\label{eq:Sthermo}
\eeq
%
%
and perfectly matches with numerics as can be seen in Fig.~\ref{fig:GSvsh} (right).

As previously, one can expand ${\mathcal S}$ in the vicinity of the critical point to get the relation
%
%
\beqa
{\mathcal S}(\tau,\gamma,h) &=& -\frac{1}{4} \ln (h-1) + \frac{1}{4}  \ln(1-\gamma) + \frac{1}{2} \ln [\tau(1-\tau)] \nonumber \\
&& -\ln 2 + {\mathcal O}\big[(h-1)^{1/4}\big] .
\label{offc}
\eeqa
%
%

Remarkably, the leading terms in the above expression are identical to those found for the entanglement entropy \cite{Barthel06_2}, the only difference occuring in the subleading corrections, which are 
${\mathcal O}\left[(h-1)^{1/4}\right]$ for ${\mathcal S}$ and 
${\mathcal O}\left[(h-1)^{1/2}\right]$ for ${\mathcal E}$ (and ${\mathcal G}$).

%
%
\emph{Finite-size behavior at the critical point ---}
%
%
Using the scaling hypothesis discussed in Refs.~\cite{Barthel06_2,Dusuel05_2}, one can further obtain the behavior of ${\mathcal G}$ and ${\mathcal S }$ at the critical point. 
For the global geometric entanglement, one gets
%
%
\beq
{\mathcal G}(\gamma,h=1) \sim \frac{1}{6} \ln N + \frac{1}{6} \ln(1-\gamma).
\label{eq:GvsN}
\eeq
%
%
Note that in the present case, the density of global geometric entanglement ${\mathcal G}/N$ vanishes in the thermodynamic limit contrary to 1D systems  where it remains finite \cite{Botero08,Orus08_1}.  
For the single-copy entanglement, one similarly obtains
%
%
\beq
{\mathcal S}(\tau,\gamma,h=1) \sim \frac{1}{6} \ln N + \frac{1}{6} \ln(1-\gamma) + \frac{1}{2} \ln[\tau (1-\tau)] .
\label{eq:SvsN}
\eeq
%
%
Here again, in the large $N$ limit, one recovers the same finite-size behavior as for the entropy so that we can finally formulate the central result of this paper~: \emph {in the critical region of the LMG model, one has} 
\beq
{\mathcal G} \sim {\mathcal S} \sim {\mathcal E} . 
\label{relation}
\eeq
%
%
\begin{figure}[t]
  \includegraphics[width=0.9\columnwidth]{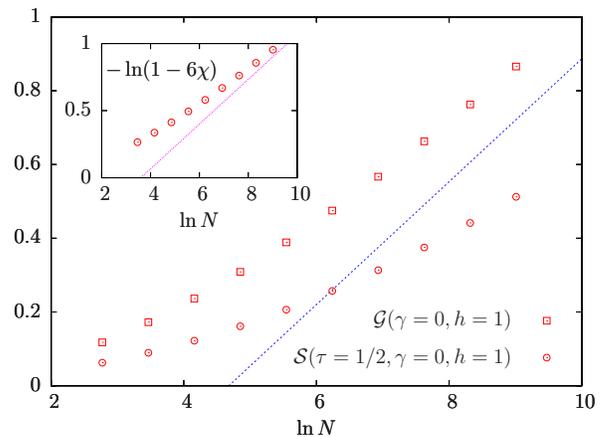}
  \caption{(color online).
Behavior of the ${\mathcal G}$ and ${\mathcal S}$ at the critical point as a
function of $N$. The blue dotted line has a slope $1/6$, as predicted by the
scaling hypothesis. The inset shows the behavior of the coefficient $\chi$ (see
text) as a funcion of $N$. The dotted line in the inset also has a slope $1/6$.}
  \label{fig:GSvsN}
\end{figure}
%
%
To check the scaling laws (\ref{eq:GvsN}) and (\ref{eq:SvsN}), we computed
numerically the behavior of ${\mathcal G}$ and ${\mathcal S}$ at $h=1$ and
$\gamma=0$ (and $\tau=1/2$ for $\mathcal{S}$) as a
function of the system size, in the range $N=16$ to $N=8192$.
Results are shown in Fig.~\ref{fig:GSvsN}, where the two straight lines have
slope $1/6$. For $\mathcal{G}$, the asymptotic regime is nearly already reached
for the maximum size $N=8192$ we used, which confirms (\ref{eq:GvsN}).
This is not the case for $\mathcal{S}$, because subleading corrections are more
important.
In order to quantify this, we have fitted $\mathcal{S}$ with the law
$\mathcal{S}=A+\chi\ln N$ in the vicinity of each value of $N$. The inset in
Fig.~\ref{fig:GSvsN} shows that $\chi\sim 1/6+B N^{-1/6}$. Thus, $\chi$ indeed
converges towards the expected value of $1/6$. This convergence is however very
slow, which explains why the asymptotic value is not yet reached even for sizes
as large as $N=8192$. 

%
%
\emph{Comparison with one-dimensional systems ---}
%
%
As said in the introduction, the above result does not hold in 1D systems. Indeed, let us call $\xi$ the correlation length of a 1D system and $L$ the size of a given block. Let us also define a quantity $l$ such that $l = L$ at criticality and $l = \xi$ away from criticality and in the regime $L \gg \xi$. In critical regions of 1D systems that are described by an underlying conformal field theory with central charge $c$, one has that \cite{Vidal03_1,Latorre04_1,Its05,Orus06_1,Botero08,Orus08_1}
%
%
\beq
\frac{1}{N}{\mathcal G}(l) \sim \frac{1}{2}{{\mathcal S}}(l) \sim \frac{1}{4} {\mathcal E}(l),
\label{onedim}
\eeq
%
%
where  ${\mathcal E}(l) \sim (c/3) \ln l$, and $N \rightarrow \infty$ is the number of blocks of size $L$ that define each party in the case of the geometric entanglement \cite{Botero08, Orus08_1}. Therefore, in 1D one has that ${\mathcal S}(l) \sim (c/6) \ln l$ and ${\mathcal G}(l)/N  \sim (c/12) \ln l$. Notice the difference between Eq.~(\ref{onedim}) and our result for the LMG model.  The appearance of the factors $1/2$ and $1/4$ in front of the single-copy entanglement and the entanglement entropy in Eq.~(\ref{onedim}) seems to be endowed with the (assumed) conformal structure at the core of the critical points in 1D. In the case of the LMG model, this conformal structure is no longer present and Eq.~(\ref{relation}) holds instead of Eq.~(\ref{onedim}). 

%
%
\emph{Discussion ---}
%
%
Our results have a clearcut interpretation from the perspective of quantum information. 
More precisely, Eq.~(\ref{relation}) establishes that the ground state of the LMG model is equally suited as a resource for two different tasks, namely, the concentration of entanglement by local operations \cite{Bennett96_1} from $(i)$ infinitely-many copies of the system (quantified by ${\mathcal E}$), and from $(ii)$ just one system (quantified by ${\mathcal S}$). 
As shown in this paper, the capability to perform such tasks is also equivalent to the fidelity between the ground state of the system and the closest separable state of all the spins (quantified by ${\mathcal G}$). 
Our results also suggest that the relationship between entanglement measures depends on the dimension of the quantum system considered. An interesting issue would be to analyze these measures in exactly solvable systems such as, for instance, the celebrated Toric Code model \cite{Kitaev03}. However, note that if the definition of ${\mathcal G}$ may be non ambiguous in any dimension, it is not the case for other measures.

\acknowledgments

R.  Or\'us acknowledges financial support from the Australian Research Council in the form of an APD Fellowship, and from The University of Queensland in the form of an Early Career Researcher grant. 


\end{document}